\documentclass[10pt, byrevtex,pre,aps,showkeys,showpacs,twocolumn,floatfix,a4paper]{revtex4}
\usepackage[dvips]{graphicx}
\usepackage{amsmath}
\usepackage{amsfonts}
\usepackage{amssymb}
\usepackage{times}

\begin{document}

\title{A toy model of faith-based systems evolution}

\author{Suzanne Sadedin}
\email{chiriyu@nosubstancesoftware.com}
\affiliation{
Department of Biological Sciences,\\ 
Monash University,\\
Victoria 3800, Australia.}

\author{Bart{\l}omiej Dybiec\footnote{Corresponding author.}}
\email{bartek@th.if.uj.edu.pl}
\affiliation{Marian~Smoluchowski Institute of Physics,\\
Jagellonian University,\\
Reymonta~4, 30--059~Krak\'ow, Poland}

\author{Gerard Briscoe}
\email{gerard@mle.media.mit.edu}
\affiliation{Dynamic Interactions Group,\\
Media Lab Europe,\\
Sugar House Lane, Bellevue, Dublin 8, Ireland}

\date{October 16, 2002}

\begin{abstract}
A simple agent-based model (ABM) of the evolution of faith-based systems (FBS)
in human social networks is presented. 
In the model, each agent subscribes to a single FBS, and may   
be converted to share a different agent's FBS during social interactions. 
FBSs and agents each possess heritable
quantitative traits that affect the probability of transmission of FBSs. 
The influence of social 
network conditions on the intermediate and final macroscopic states is examined.
\end{abstract}



\maketitle

%
% INTRODUCTION
%
\section{Introduction}

Human societies tend to hold sets of shared and interrelated beliefs that 
are not directly or easily accessible to proof or disproof. Such sets of 
beliefs, which will be called Faith-Based Systems (FBSs), may include, for 
example, cultural preferences, personal worldview, 
and belief or disbelief in one or more supreme beings. FBSs are often highly
detailed and tend to be shared
within local communities, but are highly diverse among global communities.

Dawkins \cite{dawkins} suggested that the evolution and spread of FBSs and other 
cultural information (memes) may be subject to natural selection, noting 
that both parasites and memes are transmissible between hosts and show 
heritable variation, creating the potential for competition for survival and 
reproduction. While the meme hypothesis has received considerable 
theoretical attention, there have been few attempts to explicitly model it. 

Previous models of social transmission \cite{holyst,kacperski,plewczynski,sznajd} 
have examined mainly two-state  systems; in such models, agent opinions are 
considered to be transmissible Boolean states. Such Boolean state models do not allow for the evolutionary 
potential of memetic transmission, which may influence final state distribution, 
or for the potential for a large number of FBSs to exist simultaneously in a single social system. 
In contrast, FBSs can be modelled as continuous states on a binary vector, allowing for memetic evolution
among populations of varying FBSs.

In this paper an agent-based model of the evolution and spread of 
FBSs in a two-dimensional lattice is presented. Agent-based models are useful for 
examining how local properties of individual units interact to 
produce macroscopic properties of systems, and have been applied in 
the analysis of social opinion formation \cite{holyst,kacperski,plewczynski,sznajd}, cultural interactions \cite{klemm},  
epidemiology and population dynamics \cite{sznajd2}. 
The current model explores the development of FBSs with 
evolutionarily relevant properties (faith, proselytism and mutability), in simple, 
individually varying agents. Mortality and fecundity probability curves for 
agents are applied based on human population data from the 1990s \cite{internet}. Agents 
have heritable personality characteristics that affect their susceptibility 
to conversion and their likelihood of creating a new FBS; this provides a 
non-uniform spatial environment for the evolution of FBSs. 
Characteristics of agents modelled are (cf.~Tab.~\ref{agenttable}) resistance ($R$) (the tendency to resist conversion 
to a different FBS), activity ($A$) (the tendency to participate in transmission 
of FBSs), charisma ($C$) (ability to convert other agents to their current FBS), 
and imagination ($I$) (the tendency to invent a new FBS based on the existing 
one).

The influence of two factors on the distribution and characteristics of
FBSs is examined. The frequency of social interactions of agents, where social interactions offer the potential 
for conversion between beliefs, seems likely to influence the spread and survival of FBSs.
When social interactions are less frequent, pressure for conversion between FBSs is likely 
to be smaller, but social support for existing FBSs is also decreased. If social support plays
an important role in preventing conversion, extinction of unsuccessful FBSs may occur more quickly when social 
interactions are less frequent. This could lead to decreased numbers of FBSs, each on average having a 
larger number of adherents. 
Additionally, the effect of varying total social network size on FBS evolution is explored. 
Larger total social network size may increase the influence of the social interaction condition, and 
may allow a larger number of coexisting FBSs and older FBSs (due to lower extinction rates in larger
grids).

%
% MODEL
%
\section{Model}

\begin{figure}[!h]
\begin{center}
\includegraphics[angle=0, width=6cm, height=4cm]{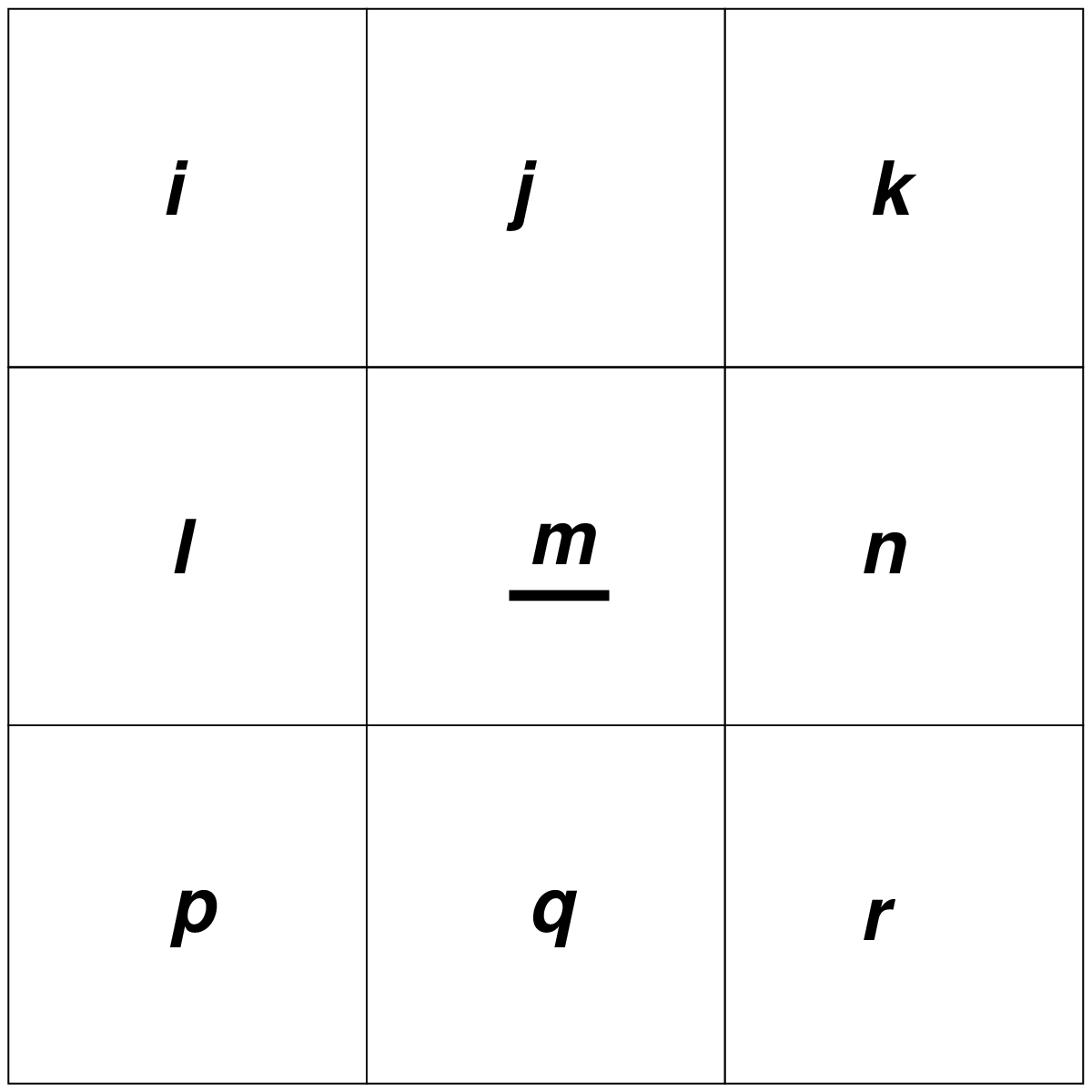}
\caption{Sample agent $m$ with its $3\times 3$ environment.
1, 4 or 8 interacting agents (social neighbors) are randomly chosen from $3\times3$ geometrical environment.}
\label{modelpl}
\end{center}
\end{figure}

The model initially consists of $N=10\times 10$ or $N=20\times 20$ agents
which are located on a regular rectangular two dimensional lattice with 
free boundary conditions.
Each agent, occupying a single site on the grid,  
interacts with up to eight social neighbors which are randomly chosen from nearest geometric neighbors; 
agents located at the border of the grid have fewer neighbors. 
(cf.~agent k on Fig.~\ref{modelpl}).
For comments on other possible grids see~\cite{flasche}.

%
% GRID PLOT
%

Within the model, social space is created by a regular rectangular lattice  and social distance is equal to 
one for the nearest interacting neighbors and infinity for others, i.e. there are no 
long range interactions (in strict meaning). 
This may be somewhat analogous to social 
transmission in pre-literate human societies, but is less applicable to 
modern societies.

%
% AGENT TABLE
%
\begin{table}[!h]
\begin{center}
\begin{tabular}{l|c|l}
\hline
\multicolumn{3}{c}{\textbf{AGENT}}\\ \hline\hline
Believe & $B$ & FBS accepted by a given agent\\ \hline
Activity & $A$ & Probability of participating  in social interactions\\
Charisma & $C$ & Ability to convert other agents\\
Imagination & $I$ & Ability to create a new FBS \\
Resistance & $R$ & Resistance to conversion by other agents \\
\end{tabular}
\end{center}
\caption{Characteristics of agents.}
\label{agenttable}
\end{table}

Every agent is characterized by a set of internal parameters which are 
constant over the life time of an agent
(cf.~Tab.~\ref{agenttable}). 
Features of agents provide a non uniform environment for the evolution and spread of FBSs.

Every FBS is characterized by a set of external parameters which are 
constant over time (cf.~Tab.~\ref{fbstable}). Since a large number of agents may
share the same FBS, a single parameter set can be widespread and sometimes dominant
throughout a grid.

%
% FBS TABLE
%
\begin{table}[!h]
\begin{center}
\begin{tabular}{l|c|l}
\hline 
\multicolumn{3}{c}{\textbf{FAITH BASED SYSTEM (FBS)}} \\ \hline\hline
Faith & $F$	 & Resistance to conversion pressure  from\\
& & other FBSs\\
Proselytism & $P$ & Strength of conversion pressure against\\
& & other FBSs\\
Mutability & $M$ & Probability of mutation  of a given FBS\\
\end{tabular}
\end{center}
\caption{Characteristics of FBSs.}
\label{fbstable}
\end{table}

All characteristic parameters of FBSs and individuals are randomly generated from 
a restricted Gaussian distribution with mean $\mu=10$ and standard deviation $\sigma=3$ and converted into integers.
The above restrictions are related only to the initial state distribution;
over time values of parameters can change due to mutation processes.

The parameters in the model (Tab.~\ref{agenttable} and~\ref{fbstable}) 
reflect a limited selection of features of real agents and FBSs. These parameters were chosen
to express some of the most influential properties of real agents and FBSs, while
limiting the parameter space explored.

Agent $m$ receives support $\mathcal{S}$ from social neighbors (surrounding interacting agents) 
(cf.~Fig.~\ref{modelpl}) which accept the same FBS $B(m)$
\begin{equation}
\mathcal{S}(B(m))  = 
\sum\limits_{s\in\{i,j,k,l,n,p,q,r\}}\delta_{B(m),B(s)}C(s)P(B(m)).
\label{suport}
\end{equation}
where $\delta_{i,j}$ is a Kronecker delta function,  $P(B(m))$ is proselytism of the FBS accepted by the agent $m$ 
and all other symbols are described in Tab.~\ref{agenttable} and~\ref{fbstable}.
The agent $m$ is also subjected to the environmental pressure
$\mathcal{P}$ 
(cf.~Fig.~\ref{modelpl})
which is caused by surrounding interacting individuals  having different FBS,
e.g. pressure caused by agent $k$ accepting FBS $B(k)$ ($B(k)\neq B(m)$) 
\begin{equation}
\mathcal{P}(B(k))  = 
\sum\limits_{s\in\{i,j,k,l,n,p,q,r\}}\delta_{B(k),B(s)}C(s)P(B(k)).
\label{pressure}
\end{equation}
The sum in eq.~(\ref{suport}) and~(\ref{pressure}) is performed over $s\in\{i,j,k,l,n,p,q,r\}$ i.e. $s$ 
takes 1, 4 or all possible values from the set $\{i,j,k,l,n,p,q,r\}$. 
The $\delta_{i,j}$ in eq.~(\ref{suport}) and~(\ref{pressure}) with appropriate indices 
allow to choose agents sharing the same FBS, which in particular may be different than accepted by the agent $m$.
Support $\mathcal{S}$ and pressure $\mathcal{P}$  (eq.~\ref{suport} and~\ref{pressure}) are weighted sums of social neighbors' charisma
as a weight the proselytism of FBSs are taken.

Every agent is characterized by total self resistance $\mathcal{R}$
\begin{equation}
\mathcal{R}(m)=R(m)F(B(m)).
\end{equation}
Total self resistance is calculated as a product of an agent's resistance
$R(m)$  and faith of an agent's FBS $F(B(m))$.

Social interactions between agents cause the agent $m$ accepting FBS $B(m)$ to be converted into the FBS $B(k)$ of 
an agent $k$  if
\begin{equation}
\mathcal{P}(B(k))>\mbox{max}\left[\mathcal{S}(B(m)),\;\mathcal{R}(m)\right].
\end{equation}
Transmission of FBSs can be interpreted in terms of epidemiology as infection, 
in that an FBS depends on a host agent, and can be transmitted between agents via specific forms of contact.
In our model, transmission is limited both by the individual resistance of an agent, 
and by the presence of social support for the agent's current FBS. This attempts to reflect the interaction of personality 
and environment in social transmission; for some individuals, personal characteristics are the main factors preventing 
conversion, while for others, social support is more important. 

New FBSs are created randomly with probability dependent on the imagination 
of the agent and the mutability of its current FBS, 
e.g. agent $m$ can create new FBS with probability $\propto I(m)M(B(m))$.
The new FBS has the same characteristics as the old FBS, with a random increase or decrease by 
1 unit to any one of its characteristic values (proselytism, faith or mutability). 
This mechanism introduces an evolutionary principle, which allows directed change in FBS characteristics over time.

The probability that a given agent (e.g. $m$) is going to participate in social 
interactions during a given iteration of the model is $\propto A(m)$.
Additionally, during each iteration, agents can move to a randomly selected adjacent cell, provided it is empty.

Space occupied by agents is limited to the total area of the grid, but the number of agents can vary 
over time due to mortality and fecundity of individuals.
The agents can reproduce, producing a new agent with similar personality 
characteristics in any adjacent empty cell. The probability of reproduction 
depends on the age of the parent agent
and is proportional to $121-(t-25)^2 $ with $t\in [14,\;36].$
The agents die at random according to a probability curve based on data for 
Australian, English, Polish and US populations during the 1990s \cite{internet}.
Conditional probability to die at given age $t$ is proportional to $(t-12)^2.$
These phenomenological relations reflect the shape of real life fecundity and mortality curves.
More details about the phenomenological theory of mortality and aging can be found in \cite{azbel}.

In the model the agents are updated in a random order.
This kind of updating corresponds to local social interactions which, as in reality, occur non-synchronously 
within small social groups.
Social interactions are governed by a deterministic rule applied to randomly chosen agents.
These dynamics create a modified version of social impact theory \cite{nowak}.

%
% RESULTS
%
\section{Results}

\begin{figure}[!h]
\begin{center}
\includegraphics[angle=0, width=8.5cm, height=12cm]{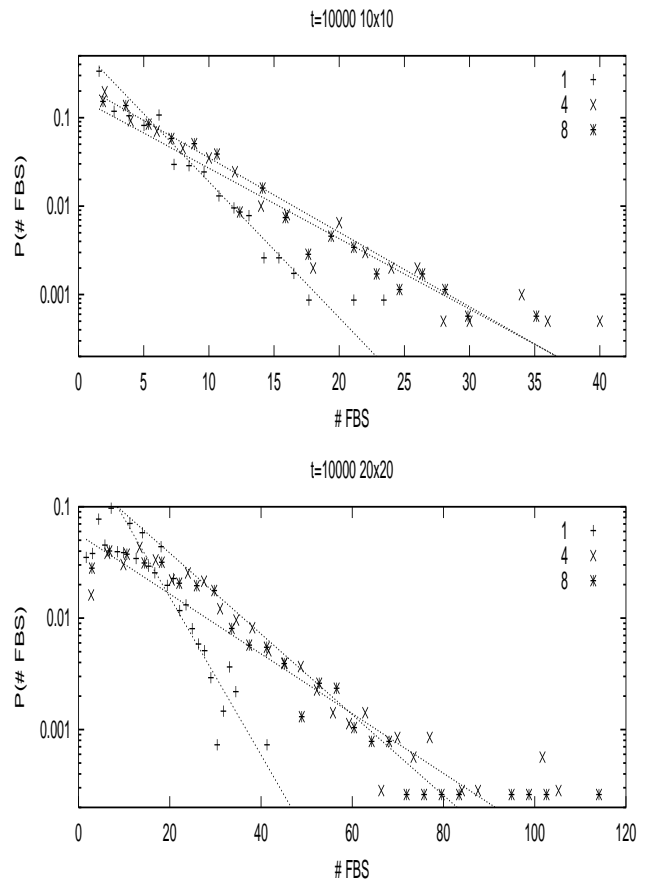}
\caption{Probability distributions (semi-logarithmic scale) of number of FBSs at $t=10000$, 
grid size $10\times10$ (left panel) and $20\times20$ (right panel), 
for a given number of interacting agents.}
\label{nr}
\end{center}
\end{figure}

\begin{figure}[!h]
\begin{center}
\includegraphics[angle=0, width=8.5cm, height=12cm]{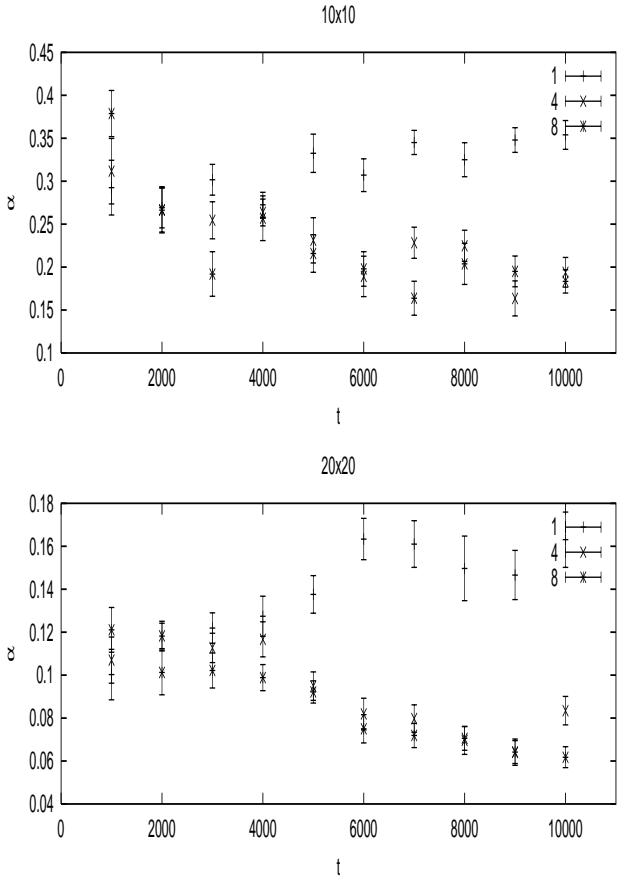}
\caption{Time behavior of characteristic exponents $\alpha$ (cf.~eq.~\ref{alpha}) for number of FBSs distributions, 
grid size $10\times10$ (left panel) and $20\times20$ (right panel), 
for a given number of interacting agents.}
\label{nrexp}
\end{center}
\end{figure}

\begin{figure}[!h]
\begin{center}
\includegraphics[angle=0, width=8.5cm, height=12cm]{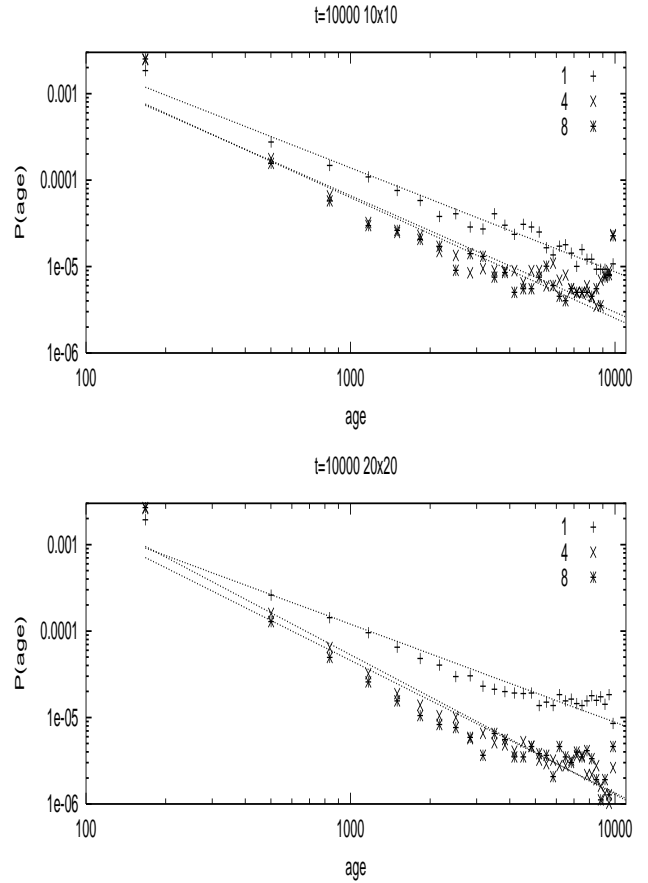}
\caption{Probability distributions  (logarithmic scale) 
of age of FBSs at $t=10000$, 
grid size $10\times10$ (left panel) and $20\times20$ (right panel),
for a given number of interacting agents.}
\label{age}
\end{center}
\end{figure}

\begin{figure}[!h]
\begin{center}
\includegraphics[angle=0, width=8.5cm, height=12cm]{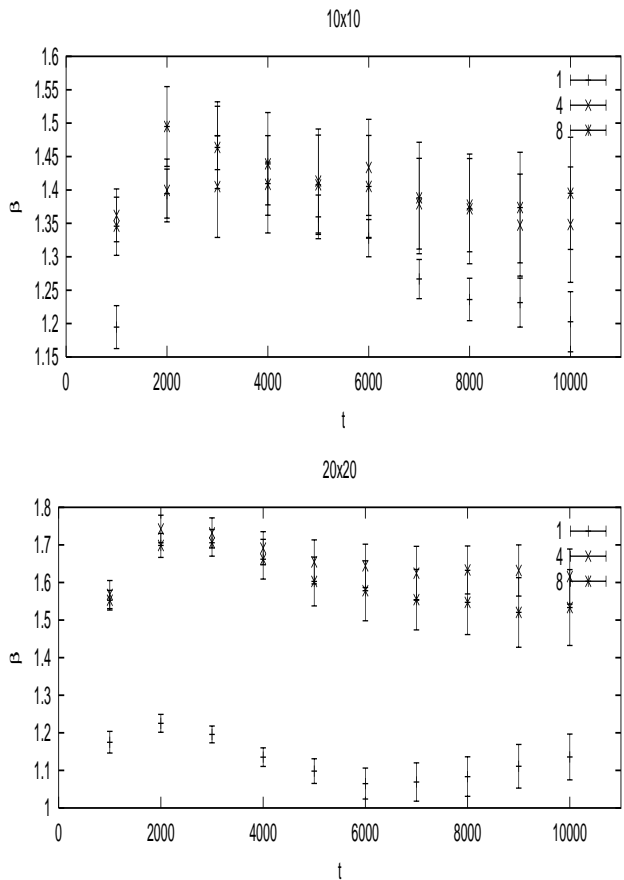}
\caption{Time behavior of characteristic exponents $\beta$ (cf.~eq.~\ref{beta})  for age of FBSs distributions, 
grid size $10\times10$ (left panel) and $20\times20$ (right panel), 
for a given number of interacting agents.}
\label{ageexp}
\end{center}
\end{figure}

\begin{figure}[!h]
\begin{center}
\includegraphics[angle=0, width=8.5cm, height=12cm]{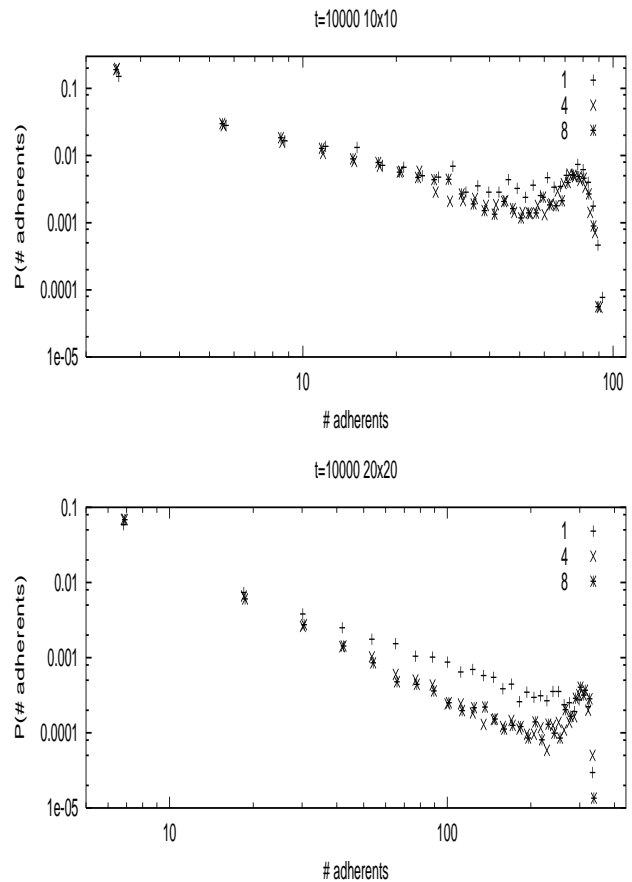}
\caption{Probability distributions (logarithmic scale) 
of number of adherents of FBSs at $t=10000$, 
grid size $10\times10$ (left panel) and $20\times20$ (right panel),
for a given number of interacting agents.}
\label{size}
\end{center}
\end{figure}

Macroscopic features of the intermediate and final state 
for different lattice sizes ($10\times 10$ and $20 \times 20$) 
and different numbers of interactions per agent per timestep (1, 4, 8 agent interactions) 
(cf.~Fig.~\ref{modelpl}) were analyzed.
Data were collected every 1000 iterations until the final state was reached (state after 10000 iterations).
To obtain sufficient statistics, around 1000 simulations were 
performed for each model considered. Probability density functions (PDFs) of number 
of FBSs (Fig.~\ref{nr}), age of FBSs (Fig.~\ref{age})
and numbers of adherents (size of FBSs)  (Fig.~\ref{size}) are shown.
Additionally, time behavior of these variables (Fig.~\ref{nrexp} and~\ref{ageexp}) and correlations between FBSs' characteristics were examined.

PDFs of a number of FBSs  seem to have exponential tails (cf.~semi-log Fig.~\ref{nr}), i.e.
\begin{equation} 
\mbox{P(\#FBS)}\propto\exp(-\alpha \mbox{\#FBS}).
\label{alpha}
\end{equation}
There is a high probability of 
a small number of FBSs, and a small probability of a large number of FBSs co-existing in a single system. 
Values of exponents $\alpha$ for the lattice $10\times10$ are higher than for the grid $20\times20$.
Time behavior of exponents $\alpha$ (with error bars) is presented in Fig.~\ref{nrexp}.
For $t>4000$ there is a difference between values of exponents $\alpha$ for binary, 1 to 1, interaction (higher exponent) 
and group, 4 or 8 to 1, interaction (lower exponent), for both lattice sizes.
Values of exponents $\alpha$ for 4 or 8 interacting agents are similar.

%
% # FBS
%

%
% # FBS exponents
%

At any point in time it is possible to find an FBS as old as the system.
There is also a significant probability of finding young and very young FBSs.
The age distribution expresses a power law (cf.~log-log plot Fig.~\ref{age}), i.e.
\begin{equation}
\mbox{P(age)}\propto \mbox{age}^{-\beta}.
\label{beta}
\end{equation}
Small deviations from a straight line, which are visible on the log-log plot 
are caused by the finite world size and limited size of random sample.
The order of magnitude is affected by the age of the model.
Time behavior of exponents $\beta$ (with error bars) is presented in Fig.~\ref{ageexp}.
Values of exponents $\beta$ for every type of interaction  depend on the size of the lattice, 
but do not change dramatically over time (cf.~Fig.~\ref{ageexp}).
For binary interactions, values of exponents $\beta$ are smaller than for group interactions.
Interactions with 4 or 8 agents seem to be equally efficient, i.e. values of exponents are similar.

%
% age
%

%
% age exponent
%

Each agent occupies a single cell on the lattice, i.e. the size of FBS is the number of adherents.
Distributions of size are shown in the Fig.~\ref{size}.
In the log-log plot the linear decay and local maxima of PDFs for both lattice sizes are visible.
For the $20\times20$ grid the linear part of the PDF is steeper than for the $10\times10$ grid.
For the bigger world size, differences in slope for binary and group interactions are observable.
The differences for the small world size are not so apparent.
Local maxima of PDFs which are visible (cf.~Fig.~\ref{size}) correspond to the situation 
when a large cluster of a single FBS occupies almost the whole lattice.
The size of a macroscopic cluster is approximately $70-80\%$ of the whole lattice.
Local maxima of PDFs are most visible for the $20\times20$ lattice with 4 or 8 agents interaction
and are not visible for binary interactions on the same lattice.
This phenomenon can be explained by a kind of ``non-local'' 
character of a social interaction with 4 or 8 social neighbors.
Despite the fact that in the interaction only some of nearest geometrical neighbors participate,
the interacting agent is more linked with its environment.
The larger number of links provides a kind of social connection, 
which introduces effects similar to long range interactions.
This is in agreement with the theory of social distance, 
which in general can be different from geometrical distance.
For lattice $10\times10$ the local maxima of FBSs' size distribution
are visible for all interaction conditions.
This is probably a consequence of the finite world size.

%
% # adherents
%

Effects of the number of interactions per agent are visible, manifested as different values of
exponents for systems with only 1 agent interaction 
compared with systems with 4 or 8 agent interactions
(Figs.~\ref{nr},~\ref{nrexp} and~\ref{age},~\ref{ageexp}).
Binary interactions seem to be more efficient in terms of FBS transmission.
This efficiency is manifested by a smaller number of FBSs existing in the 1 to 1 interaction condition, 
in comparison with the group interaction conditions. Additionally, FBSs in the 1 interaction condition 
are older and have more adherents. The phenomenon of higher efficiency can be explained by the lack 
of support for agents participating in binary interactions.
Binary interactions might be seen as corresponding to informed discussions in real social interactions, 
which provide an efficient way of convincing people by largely avoiding negotiation of opinions.

Initially all characteristics were generated as independent and identically distributed random variables,
according to the restricted Gaussian distribution.
Over time the mechanisms of conversion and mutation included in the model introduce correlations.
Positive correlations (0.58--0.80) between the age of an FBS and the number of adherents can be observed.
Within the model, it seems that even highly successful FBSs need time to become widespread. 
The variations of correlation value are caused by mortality and conversion processes.
The absolute values of other correlations are smaller.
In particular, small negative correlations (-0.17 --  -0.08) between mutability and the age of FBSs, and mutability 
and the size of FBS were observed. As mutability increased, both the age and size of FBSs decreased; this is attributed to
the increased tendency for agents holding high-mutability FBSs to invent new FBSs.
Smaller negative correlations also occurred between proselytism and age, and proselytism and size.
Because of the generally low level of correlation, it is hard to draw any detailed conclusion from these findings.
What is intriguing, however, is an unexpected sign of correlation between age and proselytism, and size and proselytism.
This effect may be caused by the limited size of our random samples or the short time of simulation.
It is also necessary to emphasize that if random variables are independent they are uncorrelated.
The inverse theorem is not necessarily true.

%
% SUMMARY
%
\section{Summary}

A simple agent-based model of the formation and evolution of FBSs has been 
presented. 
The influence of the lattice size and frequency of 
social encounters on macroscopic output have been examined in detail.

In the model, the age distributions of FBSs express
power laws (Fig.~\ref{age}). Larger social networks, 
as reflected in larger lattices, have larger numbers of coexisting FBSs; 
on average, FBSs in such systems are younger and have more adherents. 
At low levels of social 
encounters,  FBSs are, in general, older and have more adherents, than at higher levels of social encounters. 
A possible 
explanation of this pattern is that the single-encounter condition prevented agents from receiving support for
their current FBS from adjacent interacting agents of the same FBS; this could cause a higher incidence of conversion 
leading to more rapid extinction of unsuccessful FBSs.

On the distributions of the FBSs' size (cf.~Fig.~\ref{size}) local maxima of PDFs are visible.
These maxima manifest the state in which almost all agents share the same FBS.
These phenomena are clearly visible for non-binary interactions which in social systems
correspond to long range interactions.

The toy model which was presented above allows for further possible extensions.
In particular, it would be interesting to examine fitness effects of FBSs on agents, 
different distributions of agent and FBS traits, and more realistic social network and agent
traits. Such extensions may enable a better critique of the utility and robustness of models
that perceive FBSs as objects of natural selection. 

%
% ACKNOWLEDGEMENTS
%
\begin{acknowledgments}
The authors thank the organizers of the Complex Systems Summer School 2002 in Budapest,
where this project was partially completed.

One of us (BD) acknowledges a stimulating influence of the visit 
at the Graduate School of Biophysics at the Niels Bohr Institute (Copenhagen, Denmark).\\
Special thanks are directed to J.~A.~Ho{\l}yst (Warsaw University of Technology, Poland) 
for a very fruitful and inspiring discussion during $15^{\mathrm{th}}$Marian 
Smoluchowski Symposium on Statistical Physics (Zakopane, Poland). 
\end{acknowledgments}

%
% BIBLIOGRAPHY
%


\begin{thebibliography}{10}

\bibitem{dawkins}
{R.~Dawkins}, {\em The Selfish Gene}
(Oxford University Press, New York 1976), pp.~189-201.

\bibitem{holyst}
{J.~A.~Ho{\l}yst, K.~Kacperski and F.~Schweitzer}, in 
{\em Annual Reviews of Computational Physics}, vol 9,
edited by D.~Stauffer
(World Scientific, Singapore 2001), pp.~253-273.

\bibitem{kacperski}
{K.~Kacperski and J.~A.~Ho{\l}yst}, Physica A \textbf{287},  631 (2000).

\bibitem{plewczynski}
{D.~Plewczy\'nski}, Physica A \textbf{261},  608 (1998).

\bibitem{sznajd}
{K.~Sznajd-Weron and J.~Sznajd}, Int. J. Mod. Phys. C \textbf{11},  1157 (2000).

\bibitem{klemm}
{K.~Klemm, V.~M.~Eguiluz, R.~Toral and M.~San~Miguel},
{\em Global culture: A noise induced transition in finite systems} 
(arXiv:cond-mat/0205188).

\bibitem{sznajd2}
{K.~Sznajd-Weron and A.~P\c ekalski}, Physica A \textbf{274},  91 (1999).

\bibitem{internet}
For example population data are available at: 
\url{http://www3.who.int/whosis/}, \url{http://www.census.gov/}, \url{http://esa.un.org/unpp/}.

\bibitem{flasche}
{A. Flache and R. Hegselmann}, Journal of Artificial Societies and Social Simulation 
(http://www.soc.surrey.ac.uk/jasss)  \textbf{4}(4) (2001). 

\bibitem{azbel}
{M.~Ya.~Azbel}, Physica A \textbf{249},  472 (1998).

\bibitem{nowak}
{M.~Lewenstein, A.~Nowak and B.~Latan\'e}, Phys. Rev. A \textbf{45},  763 (1992).

\end{thebibliography}
\end{document}